\documentclass[%
preprint,
nobibnotes,
longbibliography,
 amsmath,amssymb,
]{revtex4-1}
\usepackage{dcolumn}% Align table columns on decimal point
\usepackage{bm}% bold math
\usepackage[dvips]{graphicx}
\usepackage{mathrsfs}
\usepackage{amsfonts}
\usepackage{amssymb}
\usepackage{color}
\usepackage{subfigure}
\usepackage{amsmath,latexsym}
\newcommand{\ignore}[1]{}
\newcommand{\rev}[1]{{\color{black} #1}}

\begin{document}

\title{Theoretical analysis for flattening of a rising bubble in a Hele-Shaw cell}
\author{Xianmin Xu}%(许现民)
 \email{xmxu@lsec.cc.ac.cn}
\affiliation{ LSEC,ICMSEC,
  NCMIS, Academy of Mathematics and Systems Science, Chinese Academy of Sciences, Beijing, 100190, China%\textbackslash\textbackslash
}
\author{Masao Doi}%(土井正男)
 \email{masao.doi@buaa.edu.cn}
\author{Jiajia Zhou}%(周嘉嘉)
\email{jjzhou@buaa.edu.cn}
\affiliation{ Center of Soft Matter Physics and its Applications, Beihang University, 37 Xueyuan Road, Beijing, 100191, China
}
\author{Yana Di}%(邸亚娜)
 \email{yndi@uic.edu.hk}
\affiliation{ Institute of Mathematical Research, Beijing Normal University \& UIC;
 Division of Science and Technology, BNU-HKBU United International College,  Zhuhai, 519087, China}

%\begin{article}
\begin{abstract}
We calculate the shape and the velocity of a bubble rising in an infinitely large and closed Hele-Shaw cell
using Park and Homsy's boundary condition which accounts for the
change of the three dimensional structure in the perimeter zone.  We first formulate the problem 
in the form of a variational problem, and discuss the shape change assuming that the
bubble takes elliptic shape.  We calculate the shape and the velocity of the bubble  
as a function of the bubble size, gap distance  and the inclination angle of the cell.
We show that the bubble is flattened as it rises. This result is in agreement with experiments for
large Hele-Shaw cells.

%The moving bubble  in Hele-Shaw cell is an interesting two-phase flow 
%problem in confined geometry.
%Experiments shows that a bubble can have different shapes in steady states. 
%It is difficult to analyse the problem theoretically since three dimensional effects near
%the edge of the bubble may affect its shape. 
%We analyse the problem by using the Onsager Principle and show that the 
%Bretherton effect is essential for the flattening of the bubble. 
%The phenomena were observed
%in many experiments but have not been studied theoretically.  In addition, the effect on
%the rising velocity of the bubble of the shape changes are also studied quantitatively.
\end{abstract}
\maketitle
\section{Introduction}
Motion of a bubble moving in a Hele-Shaw cell under gravity is a classical problem first 
discussed by Taylor and Saffman in 1959\cite{taylor1959note}, 
yet there remains an unsolved problem. 
To make the discussion clear, we restrict ourselves to the problem of an isolated bubble rising 
under gravity in a closed and infinitely large Hele-Shaw cell. 
The problem is how the shape of the rising bubble is determined.

Taylor and Saffman\cite{taylor1959note} showed that the set of equations determining 
the shape and the velocity of the bubble in steady state can be solved analytically 
if the effect of surface tension is ignored. They also showed that there are infinite 
number of such solutions, and further condition is needed to determine 
the shape uniquely.  They made a conjecture which determines the unique 
solution observed in experiments, but they could not justify the physical or mathematical
origin of the conjecture.

%{\color{black}Jiajia: \it  I think McLean and Saffman first did the calculation with the surface tension (JFM, 102, 455 (1981)).}
Twenty seven years later, Tanveer\cite{tanveer1986effect} 
showed that the degeneracy of the Taylor-Saffman solution is 
removed if the surface tension is accounted for, but there still remain multiple branches 
of exact solutions\cite{tanveer1987new}. Furthermore, many other 
solutions have been found in recent years\cite{crowdy2009multiple,green2014multiple,green2017effect}.

Experimentally, the multiplicity of the solutions is puzzling. It has been observed 
that if the rising velocity $U$ is small, the bubble takes a circular shape, and with 
increasing the velocity, the bubble deforms to ellipse, and cambered ellipse\cite{eck1978bubble}. 
Kopf-Sill and Homsy\cite{kopf1988bubble} studied the bubble shape when various parameters, such as rising speed, bubble size, liquid viscosity are varied.
%shape, changing various parameters, the rising speed, bubble size, and the liquid viscosity etc.
 They have shown 
that for a bubble rising in a large cell, the bubble shape changes from circle to flattened ellipse 
(with long axis perpendicular to the moving direction), but no theory has been given to explain such shape change.

Recently, the rising bubble has been studied both theoretically and experimentally 
and also by simulation \cite{eri2011viscous,yahashi2016scaling,okumura2018viscous,
keiser2018dynamics,shukla2019film,wang2016volume,tihon2019velocity}. Theories have been given for 
the rising velocity of a bubble of given  
shape\cite{eck1978bubble,meiburg1989bubbles}, 
but no theories have been given to predict the shape of the bubble as far as we know.

The lack of the theory predicting the bubble shape is related to the fact, first shown by 
Tanveer\cite{tanveer1986effect}, that perturbative calculation cannot be performed for 
the shape change of the bubble. One expects that when a bubble starts to move, 
it changes the shape from circular to elliptic. Tanveer, however, has shown that the 
circular solution is an isolated solution which is always valid, and other solutions 
cannot be obtained by perturbation method.

There is other difficulty in calculating the bubble shape. The bubble shape we 
are talking about is the shape of the perimeter in the 2D plane parallel to the cell wall. 
However, the perimeter of the bubble in the Hele-Shaw cell is not a line, 
but a region having a length of the order of the gap thickness. 
The 3D structure of this region influences the 2D shape of the bubble\cite{kopf1988bubble,meiburg1989bubbles}.
 In the classical works of Tayler-Saffman and Tanveer, the interfacial region was regarded as a line 
across which the pressure changes discontinuously. The discontinuity in the pressure is given by 
the air/fluid surface tension times twice of the mean curvature of the interface, i.e., the average of the 
curvature in the plane perpendicular to the cell wall, and that in the plane parallel to the cell wall. 
Taylor and Saffman conducted the analysis assuming that the first curvature is dominant and is 
constant\cite{taylor1959note}. This assumption becomes equivalent to setting the surface 
tension zero in the present problem. Tanveer took into account of the effect of the second 
curvature, but this was not enough since the first curvature also changes when the interface 
is moving as it was first shown by Bretherton\citep{Bretherton}. 
Taylor and Saffman discussed this effect in their classical work\cite{saffman1958penetration,taylor1959note}, 
but did not develop a theory for it. Park and Homsy considered this effect and derived a new boundary 
condition for the perimeter\cite{park1984two}. Their boundary condition makes the problem 
non-linear and  thus difficult to handle analytically. Accordingly, their boundary 
condition has not been used in previous studies apart from in numerical simulations\cite{meiburg1989bubbles}.

In this paper, we shall calculate the deformation of a rising bubble using Park-Homsy's boundary condition. 
We take an approach different from  previous ones. We first show that the set of equations
to be solved can be derived by a minimization of certain functional for the shape change of the bubble, 
and then determine the shape assuming a elliptical shape of the bubble.  This approach is not exact, 
but it allows us to have an analytical expression for the shape and the velocity of the bubble as a 
function of various experimental parameters.  The same approach has been used in many other problems
\cite{XuXianmin2016,ManXingkun2016,DiYana2018,guo2019onset}. 

The structure of this paper is as follows. In Section II, we review the boundary condition by Park and Homsy 
and state the problem in the form of a variational problem.  In Section III, we consider the motion of a bubble in a 
Hele-Shaw cell and derive a reduced model using the variational principle.
In Section IV, we analyze the reduced model and discuss the shape and the velocity of the bubble
as a function of various physical parameters.  Finally we conclude briefly in Section V. 

\section{Variational formulation}

\subsection{Basic equation}
\begin{figure}[!htp]
    \includegraphics[width=10cm]{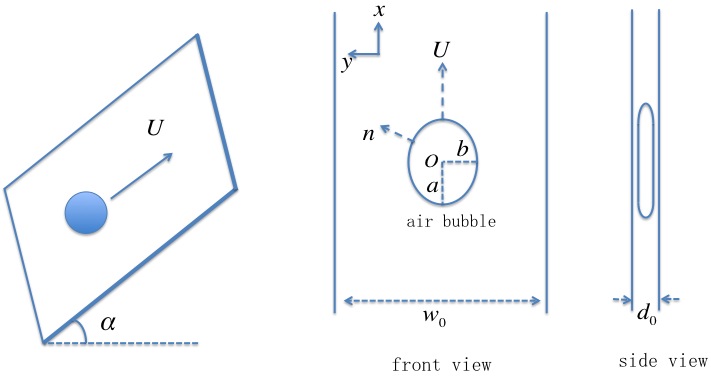}
    \vspace{-0.5cm}
    \caption{A bubble in a Hele-Shaw cell.}%n张图片共享的说明
    \label{fig:heleshaw}
\end{figure}

We consider a very large Hele-Shaw cell filled with a liquid of viscosity $\mu$ tilted against the
horizontal plane with angle $\alpha$ (see Fig,~\ref{fig:heleshaw}).
Inside the liquid there is a small air bubble which rises with certain velocity $U$ due to gravity. 
The gap distance  $d_0$ of the Hele-Shaw cell is assumed to be
much smaller than the bubble size, and the capillary length $\sqrt{\gamma/\rho g}$ (where 
$\rho$ and $\gamma$ are the density and the surface tension of the liquid respectively). Therefore
the bubble takes a pancake shape of thickness $d_0$ between the cell wall; 
the thickness of the liquid film between the bubble and the plates is ignored. 

%We consider a Hele-Shaw cell as in Figure~\ref{fig:heleshaw}. 
%The liquid is confined between two parallel plates. The size of 
%the gap between the two plates is $d_0$.
%Inside the liquid there is an air bubble.  

We take $x$-$y$ coordinate in the plane of the Hele-Shaw cell, with the y axis 
being in the horizontal plane.  Let $\bm{u}(\bm{r})$ be the depth-average 2D velocity
of the fluid at point $\bm{r}$, $\bm{u}$ satisfies the Darcy equation,
\begin{equation}\label{e:darcy}
  \bm{u}=-k(\nabla p +\rho g \sin\alpha \mathbf{e}_x),
\end{equation} 
where $k=\frac{d_0^2}{12\mu}$,  $p$ is the pressure and
$\mathbf{e}_x$ is the unit vector along the $x$ axis.
% \rev{$\phi=k(p+\rho g \sin \alpha x)$ is called velocity potential since $u= - \nabla \phi$.}
The velocity $\bm{u}$ satisfies the incompressible condition,
\begin{equation}\label{e:incompressible}
\nabla\cdot \bm{u}=0.
\end{equation}
Equations (\ref{e:darcy}) and (\ref{e:incompressible}) gives the Laplace equation for $p$
\begin{equation}\label{e:laplace}
    \nabla^2  p=0.
\end{equation}
Therefore $\bm{u}$ is obtained if the boundary condition for $p$ is known. 

Park and Homsy \cite{park1984two} conducted asymptotic analysis for the problem and 
derived the following effective boundary condition for the pressure $p$(see also \cite{reinelt1987interface}),
\begin{align} 
 p%+ \rho g x \sin \alpha 
 &
       =-\frac{2\gamma}{d_0}(1+\beta Ca_n^{2/3}+\cdots)-\frac{\gamma}{R(s)}(\frac{\pi}{4}+O(Ca_n^{2/3})),
                                  \label{e:2.1}     \\
[u]&=O(Ca_n^{2/3}).
\end{align}
Here $Ca_n=\frac{\mu |\bm{u} \cdot\mathbf{n}|}{\gamma}$  
is the capillary number defined for the normal velocity  
$u_n = \bm{u} \cdot\mathbf{n}$ at the boundary of the bubble, $R(s)$ is the local 
radius of the curvature of the bubble, and
 $\beta$ is a numerical constant which is equal to $3.8$ when the interface is locally 
advancing($u_n>0$) and equal to $-1.9$ when the interface is receding($u_n<0$). 
$[u]$ is the difference between the velocity $u_n$ and the moving velocity of the bubble boundary.
Introducing the velocity $U^*= \gamma/ \mu$, the capillary number $Ca_n$ is written as
\begin{equation}
     Ca_n = \frac{|u_n|}{U^*}
\end{equation}

When $Ca_n$ is small, we can consider the leading order only, and Equ.( \ref{e:2.1}) becomes 
\begin{align}
p%+{\rho g x\sin\alpha}
&=-\frac{2\gamma}{d_0}(1+\beta Ca_n^{2/3})-\frac{\gamma}{R(s)}\frac{\pi}{4},\label{e:BndP0}\\
[u]&=0.\label{e:Bndv}
\end{align}
Here we have kept  the term $\frac{2\gamma}{d_0} Ca_n^{2/3}$ since $Ca_n^{2/3}/d_0$ 
may be comparable with ${1}/{R(s)}$ when $d_0/R(s)$ is small. 
Furthermore, the constant $\frac{2\gamma}{d_0}$ in Equ.~\eqref{e:BndP0} can be 
ignored by shifting $p$ by a constant. Therefore we finally have the following boundary conidition
\begin{equation}\label{e:BndP1}
p%+{\rho g x\sin\alpha}
=-\frac{2\gamma}{d_0}\beta Ca_n^{2/3}-\frac{\gamma}{R(s)}{\frac{\pi}{4}}.
\end{equation}
The other boundary condition far from the bubble is obtained from the condition that there is no flow:
\begin{equation}\label{e:BndP2}    \mbox{for }  |\bm{r} | \to \infty, \qquad 
   p \to - {\rho g x\sin\alpha}.
\end{equation}
Equ.~\eqref{e:laplace} and the boundary conditions \eqref{e:BndP1} and \eqref{e:BndP2}
compose the basic equations for the rising bubble problem in the Hele-Shaw cell.

 \subsection{Variational formulation}
The basic equations described above can be derived 
from a variational principle similar to the Onsager variational principle~\citep{DoiSoft}. 
We define a functional called Rayleighian $\mathcal{R}[\bm{u}(\bm{r})]$ which is a functional
of the velocity field $\bm{u}(\bm{r})$.  $\mathcal{R}[\bm{u}(\bm{r})]$ is chosen in such a way that 
the minimum condition of the functional gives the same set of equations
given in the previous subsection.  The Rayleighian $\mathcal{R}[\bm{u}(\bm{r})]$  
consists of two parts: one is the energy dissipation part $\Phi[\bm{u}(\bm{r})]$ which is related to the
energy dissipation (or entropy production) created in the system when the viscous fluid  is 
flowing with velocity field $\bm{u}(\bm{r})$, and the other part is related to the 
free energy change rate $ \dot A [\bm{u}(\bm{r})]$ when the fluid elements are moving with velocity $\bm{u}(\bm{r})$. 
\begin{equation}
     \mathcal{R}[\bm{u}(\bm{r})]=\Phi[\bm{u}(\bm{r})]+\dot{A}[\bm{u}(\bm{r})].
\end{equation}

In the present problem, the functional of the dissipation $\Phi[\bm{u}(\bm{r})]$ is given by the 
sum of two integrals, 
\begin{equation}
   \Phi[\bm{u}(\bm{r})] = \Phi_{bulk}[\bm{u}(\bm{r})] + \Phi_{Breth}[\bm{u}(\bm{r})],
\end{equation}
where
\begin{equation}
  \Phi_{bulk}[\bm{u}(\bm{r})]  =\frac{ d_0}{2k}\int_{\Omega^c} \bm{u}^2 dxdy     \label{eqn:2.4}
\end{equation}
stands for the energy dissipation in the bulk, and  
\begin{equation}                
   \Phi_{Breth}[\bm{u}(\bm{r})] = \frac{6\mu (U^*)^{1/3} }{5}\int_{\partial\Omega} \beta(u_n)  u_n^{5/3} ds
                                                                                                    \label{eqn:2.5}
\end{equation}
stands for the extra energy dissipation due to the motion of the perimeter. In Equ.~\eqref{eqn:2.4},
 $\Omega^c$ denotes the 2D region in the Hele-Shaw cell occupied by the liquid
and $\partial \Omega$ denotes the inner boundary of $\Omega^c$. The function $\beta(u_n)$ takes the value of 
$\beta_1=3.8$ when $u_n>0$ %(the region where the boundary is moving from gas phase to lqiuid phase), 
and the value $\beta_2=-1.9$ when $u_n<0$. 
We shall call $\Phi_{bulk}$ bulk dissipation, and $\Phi_{Breth}$ Bretherton dissipation.  
Detailed discussion on Equ.(\ref{eqn:2.5}) is  given in  Appendix A.

The free energy of the system is given by the sum of the gravitational energy and the surface energy
\begin{equation}\label{e:eng}
    A=\int_{\Omega^c}{\rho g d_0  x \sin \alpha } dxdy+\frac{\pi \gamma d_0}{4}\int_{\partial\Omega} ds.
\end{equation}
$\dot A$ is given by the time derivative of $A$, and is calculated as
\begin{equation}\label{e:RateEng}
   \dot{A}={\rho g d_0 \sin\alpha  } \int_{\Omega^c} \bm{u}\cdot \bm{e_x}dxdy
                           %+{\rho g d_0 \sin\alpha  }\int_{\partial\Omega} x u_n ds
                           +\frac{\pi \gamma d_0}{4}\int_{\partial\Omega} u_n \kappa ds,
\end{equation}
where $\kappa=\frac{1}{R(s)}$ is the local curvature of the boundary $\partial\Omega$.

We  minimize the Rayleighian with respect to $\bm{u}$ under the constraint  
$\nabla\cdot \bm{u}=0$. Introduce a Lagrangian multiplier $d_0 p$ and denote
\begin{equation}\label{e:ModRay}
\mathcal{R}_{p}= \mathcal{R}-d_0 \int_{\Omega^c} p \nabla \cdot u dx dy.
%= \mathcal{R}+d_0\int_{\partial\Omega}p u_nds +d_0 \int_{\Omega^c} u\cdot \nabla p  dx dy.
\end{equation}
%Direct computations for the variation of the functional $\mathcal{R}_{p}$ leads to
%\begin{align*}
%\frac{\delta \mathcal{R}_{p}}{\delta u}=\int_{\Omega^c}
%\end{align*}
By integration by part (noticing that $\mathbf{n}$ points into $\Omega^c$), 
\begin{equation*}
\mathcal{R}_{p}= \mathcal{R}+d_0\int_{\partial\Omega}p u_nds +d_0 \int_{\Omega^c} u\cdot \nabla p  dx dy.
\end{equation*}
One can easily verify that the Euler-Lagrange equation of the functional $\mathcal{R}_{p}$ 
gives the equation~\eqref{e:darcy} and the boundary condition~\eqref{e:BndP1} in the previous subsection.

The above variational formula is similar to that of the standard Onsager 
principle\cite{XuXianmin2016,ManXingkun2016,DiYana2018,guo2019onset}. 
The only difference is that the 
dissipation function is not a quadratic form with respect to $\bm{u}$.  This is due to the 
 non-quadratic term on the boundary $\partial\Omega$ arising from the Bretherton energy dissipation.
In the following, we will use the  variational formula as an approximation tool
  to study the shape changes of the rising bubble in the Hele-Shaw cell.

\section{Derivations of a reduced model for a rising bubble}
\subsection{Ansatz of the problem}
To analyse the shape changes of the bubble, we assume that the bubble is elliptic 
as shown in Fig.~\ref{fig:heleshaw}. \rev{It has been shown both theoretically 
\cite{taylor1959note} and experimentally\cite{kopf1988bubble} that the elliptic shape is
a good approximation when the deviation from the circular shape is small}
The radii of the ellipse in $x$ and $y$ directions  are $a$ and  $b$ respectively. 
The vertical velocity of the center of the bubble is $U$. 
Since the volume $V_0$ of  the bubble is estimated to be $\pi ab d_0$ (where the volume of the
liqid between the gas and the wall and that in the perimeter region is ignored), and
is constant, $b$ is given by
 $$
      b=\frac{V_0}{\pi d_0 a}.
 $$
 So there are two parameters to be determined, $a$ and $U$. In the following we will use 
the variational  principle  to derive a reduced dynamic model for them.
 
\subsection{Free energy}
The free energy of the system consists of the interface energy and 
the gravitational energy. The interface energy is given by
\begin{equation}
A_{surf} =  2\pi\gamma ab+\frac{\pi \gamma d_0 L}{2},
\end{equation}
where $L$ is the 2D contour  length of the boundary of the ellipse.
Since $\pi ab = V_0/d_0$  is constant, the time derivative of $A_{surf}$ is calculated as
\begin{equation}
\dot{A}_{surf}=\frac{\pi \gamma d_0}{2} \dot{L}.
\end{equation}
Using the approximation $L\approx \pi(\frac{3}{2}(a+b)-\sqrt{ab})$, we have
\begin{equation}
\dot{A}_{surf}=\frac{3\pi^2 \gamma d_0}{4} (1-\frac{b}{a})\dot{a}.
\end{equation}

The gravitational energy is given by $A_{grav}=-\rho g X \sin\alpha V_0$, where $X$ is the $x$ coordinate
of the center of mass of the bubble. Since $\dot{X}=U$,  the time derivative
 of the gravitational energy is written as
\begin{equation}
\dot{A}_{grav}=-\rho g   V_0 U\sin\alpha.
\end{equation}
Therefore $ \dot{A} =\dot{A}_{grav}+\dot{A}_{surf}$ is given by
\begin{equation}
\dot{A} = -\rho g V_0 U \sin\alpha +\frac{3\pi^2 \gamma d_0}{4}(1-\frac{b}{a})\dot{a}.
\end{equation}

\subsection{Energy dissipation functions.}
The energy dissipation can also be expressed in terms of $\dot a$ and $U$. 
If $\dot a$ and $U$ are given, the velocity field $\bm{u}(\bm{r})$ is calculated, and therefore
the functional $\Phi[\bm{u}(\bm{r})]$ can be written as a function $\Phi(\dot a, U)$. The
function $\Phi(\dot a, U)$ is equal to the minimum value of the functional $\Phi[\bm{u}(\bm{r})]$
for given boundary condition at $\partial \Omega$.

{\it Bulk dissipation.} 
If the origin of the coordinate system is taken at the center of the bubble, 
the boundary of the bubble  is written as
$$
x=a\cos\theta, \quad y=b\sin\theta,\qquad \theta\in(0,2\pi].
$$
 The corresponding outer normal direction is given by 
$$
      \bm{n}=\frac{1}{\sqrt{b^2\cos^2\theta+a^2\sin^2\theta}}(b\cos\theta,a\sin\theta)^T. 
$$
When the center is moving at velocity $U$ and $a$ is changing at rate $\dot a$, 
the normal velocity of the boundary $u_n = \bm{u} \cdot \bm{n}$ is calculated as
\begin{align*}
  u_n=U f_1(a/b,\theta)+\dot{a} f_{2}(a/b,\theta),
\end{align*}
where
\begin{eqnarray} 
    f_1(a/b,\theta)&=&\frac{\cos\theta}{\sqrt{\cos^2\theta+(a/b)^2\sin^2\theta}} \\
    f_2(a/b,\theta)&=& \frac{\cos 2\theta}{\sqrt{\cos^2\theta+(a/b)^2\sin^2\theta}}
\end{eqnarray}
If the velocity $u_n$ at the boundary is given, the velocity field $\bm{u}(\bm{r})$
in the bulk is given by $\bm{u} =- (1/k) \nabla \tilde p$, where $\tilde p$ is the solution of
the Laplace equation~\eqref{e:laplace} satisfying the boundary condition 
$\bm{n} \nabla \tilde p = -k u_n$ at $\partial \Omega^c$, and $\nabla \tilde p \to 0$ at infinitely far from
the bubble. The solution of this equation is written as
\begin{equation}
   \tilde p(x,y)=\frac{r_0 U}{k}\psi_1(\frac{x}{r_0},\frac{y}{r_0}) 
                   +\frac{r_0\dot{a}}{k}\psi_2(\frac{x}{r_0},\frac{y}{r_0}) ,
\end{equation}
where $\psi_i$ ($i=1,2$) is the solution of the following dimensionless equation
\begin{equation}\label{e:aux1}
\left\{\begin{array}{ll}
-\Delta \psi_i=0, &\hbox{in } \hat \Omega^c\\
\nabla \psi_i \cdot \tilde{\mathbf n} =  f_i(a/b, \theta),  & \hbox{on }\partial \hat \Omega\\
\nabla \psi_i\cdot \tilde{\mathbf n} \rightarrow 0 & \hbox{as } |\hat{\mathbf{r}}|\rightarrow \infty.
\end{array}
\right.
\end{equation}
Here the radius $r_0= \sqrt{ab}$ is taken to be the unit of length, 
and the domain $\hat{\Omega}^c$ is defined by  
$\hat{\Omega}^c:=\{({x}/{r_0},{y}/r_{0})|(x,y)\in {\Omega}^c\}$.
%(The domain $\hat{\Omega}$ and $\hat{\Omega}^c$ can be defined similarly.)

Therefore the energy dissipation function in the bulk region is computed as
\begin{align}
\Phi_{bulk}&=\frac{6\mu}{d_0}\int_{\Omega^c} (k\nabla \tilde p)^2 dx dy= \frac{6\mu}{d_0}\int_{\hat \Omega^c} | {r_0 U}\nabla\psi_1 +{r_0\dot{a}} \nabla \psi_2|^2  d\hat{x}d\hat{y} \nonumber\\
&=\frac{6\mu r_0^2}{d_0}(k_{11} U^2 +2k_{12}U\dot{a}+k_{22}\dot{a}^2),
\label{e:phi1}
\end{align}
where 
\begin{equation}
   k_{ij}=\int_{\hat{\Omega}^c} \nabla \psi_i\cdot\nabla \psi_j   d\hat{x}d\hat{y}
\end{equation}
By integration by part, the coefficients $k_{ij}$ can be written as
\begin{equation}
         k_{ij}=\int_{\partial\hat{\Omega}}  \psi_i f_j(a/b,\hat{\theta})d\hat{s},
\end{equation}
For elliptic bubble, the Laplace equation(\ref{e:aux1}) can be solved analytically  
and $k_{ij}$ is calculated analytically(see Appendix B)
\begin{equation}
         k_{11}=\frac{\pi b}{a}, \qquad  k_{12}=0, \qquad  k_{22}= \frac{\pi b}{2a}
\end{equation}

It is important to note $k_{12}$ is zero. This implies that there is no term which couples  
the translational motion and the shape change in the bulk dissipation $\Phi_{bulk}$. In other words,  
the bubble remains circular  if the Bretherton dissipation $\Phi_{Breth}$
is not considered.  The fact that  $k_{12}$  becomes zero
for ellipse can be shown by symmetry argument. Since the elliptic bubble is symmetric 
with respect to $y$ axis, the dissipation function must be even with respect to $U$, 
i.e. $\Phi_{bulk}(\dot{a},U)=\Phi_{bulk} (\dot{a},-U)$. This gives $k_{12}=0$.  

{\it Bretherton dissipation.} 
Given $u_n$, the Bretherton dissipation can be calculated straightforwardly by Equ.~\eqref{eqn:2.5}:
\begin{align}
\Phi_{Breth}&=\frac{6\mu}{5} 
     \int_{\partial\Omega } \beta(u_n) u_n^2(\frac{\mu |u_n|}{\gamma})^{-\frac{1}{3}}ds  \nonumber\\
          &=\frac{6\mu^{2/3}\gamma^{1/3}r_0}{5}\int_0^{2\pi} \beta(\theta) |f_1 U+f_2 \dot{a}|^{5/3}\sqrt{(b/r_0)^2\cos^2\theta+(a/r_0)^2\sin^2\theta}d\theta.\label{e:Phi2n}
\end{align}

\subsection{Evolution equation}
Given $\Phi=\Phi_{bulk}+\Phi_{Breth}$ and $\dot A=\dot A_{grav}+\dot A_{surf}$ 
as a function of $\dot a$ and $U$, the time evolution of the bubble is given by
\begin{equation}
 \frac{\partial \Phi}{\partial U} +\frac{\partial \dot A}{\partial U} =0,\qquad 
 \frac{\partial \Phi}{\partial \dot{a}} +\frac{\partial \dot A}{\partial \dot{a}} =0.
\end{equation}
This gives the following equation for $U$ and $\dot{a}$:
\begin{align}\frac{\rev{12}\mu r_0^2}{d_0}k_{11} U &+{2\mu^{2/3}\gamma^{1/3}b}\int_{0}^{2\pi}\beta(\theta)\frac{(f_1 U+f_2 \dot{a})f_1}{|f_1 U+f_2 \dot{a}|^{1/3}}\sqrt{\cos^2\theta+(a/b)^2\sin^2\theta}d\theta  \nonumber\\
&=\rho g V_0 \sin\alpha ,\label{e:dyn1}\\
\frac{\rev{12}\mu r_0^2}{d_0}k_{22} \dot{a}&+{2\mu^{2/3}\gamma^{1/3}b}\int_{0}^{2\pi}\beta(\theta)\frac{(f_1 U+f_2 \dot{a})f_2}
{|f_1 U+f_2 \dot{a}|^{1/3}}\sqrt{\cos^2\theta+(a/b)^2\sin^2\theta}d\theta \nonumber\\
&=-\frac{3\pi^2 \gamma d_0}{4}(1-\frac{b}{a}).\label{e:dyn2}
\end{align}
This equation can be solved for $\dot a$ and $U$, and it determines the time evolution of the bubble shape.

\begin{figure}[!htp]
 %   \begin{minipage}[t]{0.33\linewidth}%设定图片下字的宽度，在此基础尽量满足图片的长宽
 %   \centering
    \includegraphics[width=14cm]{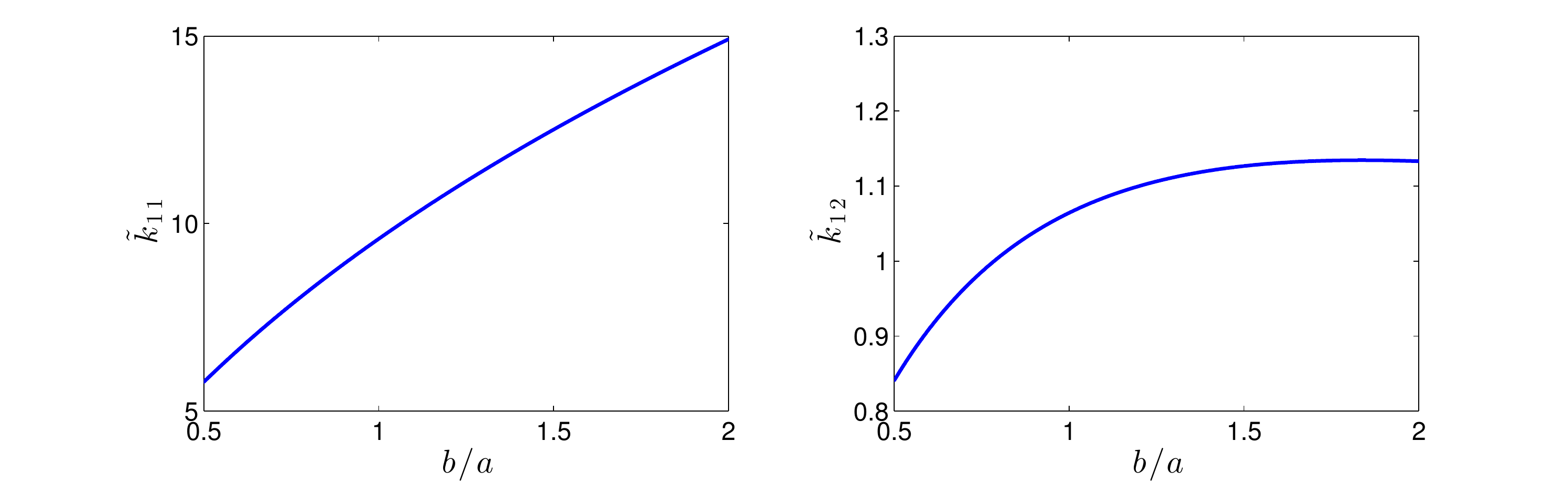}%height=4.5cm,
 %   \caption*{(a)Contact point position moves.}%加*可以去掉默认前缀，作为图片单独的说明
  %  \end{minipage}
  \vspace{-0.5cm}
    \caption{Dependence of the coefficients $\tilde{k}_{11}$ and $\tilde{k}_{12}$ on the shape parameter $S=b/a$.}%n张图片共享的说明
    \label{fig:Coef}
\end{figure}

If we are  interested only in the steady state of the bubble, we have $\dot{a}=0$. Then the equations~\eqref{e:dyn1}-\eqref{e:dyn2} are simplified to
\begin{align}
&\frac{\rev{12}\mu r_0^2}{d_0}k_{11} U +{2\mu^{2/3}\gamma^{1/3}r_0U^{2/3}}\tilde{k}_{11}=\rho g V_0 \sin\alpha ,
                    \label{e:steady1}\\
&{2\mu^{2/3}\gamma^{1/3}r_0 U^{2/3}}\tilde{k}_{12} =-\frac{3\pi^2 \gamma d_0}{4}(1-\frac{b}{a}),
                   \label{e:steady2}
\end{align}
where we have introduced two dimensionless coefficients
\begin{align}
 \tilde{k}_{11}&=\int_{0}^{2\pi}\beta(\theta)|f_1|^{5/3} \sqrt{(b/r_0)^2\cos^2\theta+(a/r_0)^2\sin^2\theta}d\theta,\label{e:tildk11}\\
 \tilde{k}_{12}&=\int_{0}^{2\pi}\beta(\theta)\frac{f_1 f_2}{|f_1|^{1/3}}\sqrt{(b/r_0)^2\cos^2\theta+(a/r_0)^2\sin^2\theta}d\theta.\label{e:tildk12}
\end{align}
Since $r_0=\sqrt{ab}$, $\tilde{k}_{11}$, and $\tilde{k}_{12}$ depends on the ratio $b/a$ only.
We call this ratio the shape parameter and denote it by $S$
\begin{equation}
     S= \frac{b}{a}.
\end{equation}
Fig.~\ref{fig:Coef} shows  $\tilde{k}_{11}$, and $\tilde{k}_{12}$ as a function of $S$. 
When $S$ changes from 0.5 to 2, $\tilde{k}_{11}$ changes significanlty, while 
$\tilde{k}_{12}$ remains almost constant (changes from 0.85 to 1.1).

\section{Results and discussions}

\subsection{Rising velocity} 
We first discuss the  rising velocity of the bubble.  The rising velocity is determined by the
balance of two forces, the gravity and the frictional force.  The gravity is 
expressed by the dimensionless number $Bo_{\alpha}$ called Bond number 
\begin{equation}
    Bo_{\alpha}=\frac{\rho g r_0^2 \sin\alpha }{\gamma}.
\end{equation}
This represents the effect of inclination angle of the cell. 
The frictional force is determined by the left hand side of Equ.~\eqref{e:steady1}. 

\rev{ The equations~\eqref{e:steady1}-\eqref{e:steady2} can be rewritten in a dimensionless form as,
\begin{align}
&\frac{12 r_0^2 S }{ d_0^2 }\frac{U}{U^*}+\frac{2r_0}{\pi d_0} \tilde{k}_{11} \Big(\frac{U}{U^*}\Big)^{2/3}=Bo_{\alpha},\label{e:steady1nonD}\\
&\frac{8 r_0}{3\pi^2 d_0}\tilde{k}_{12}  \Big(\frac{U}{U^*}\Big)^{2/3}=1-S,\label{e:steady2nonD}
\end{align}
 where $U^*={\gamma}/{\mu}$.}
If we ignore the $\tilde k_{11}$ term ( the Bretherton term) in  Equ.~\eqref{e:steady1nonD}, 
the velocity is given by 
\begin{equation}\label{e:velF1}
   U=U^* \frac{d_0^2 }{\rev{12}  r_0^2 S} Bo_{\alpha} 
     = \frac{d_0^2 \rho g \sin \alpha}{\rev{12} \mu S}.
 \end{equation}
\rev{This is exactly the velocity for small elliptic bubble given by Taylor and Saffman\cite{taylor1959note}. There 
they did not consider 
the Bretherton term and $S$ can be chosen freely. 
If the bubble is circular, the rising velocity becomes}
\begin{equation}\label{e:velF2}
   U_{circle}  = \frac{d_0^2 \rho g \sin \alpha}{\rev{12} \mu }.
 \end{equation}

\begin{figure}[!htp]
 %   \begin{minipage}[t]{0.33\linewidth}%设定图片下字的宽度，在此基础尽量满足图片的长宽
 %   \centering
    \includegraphics[width=9cm]{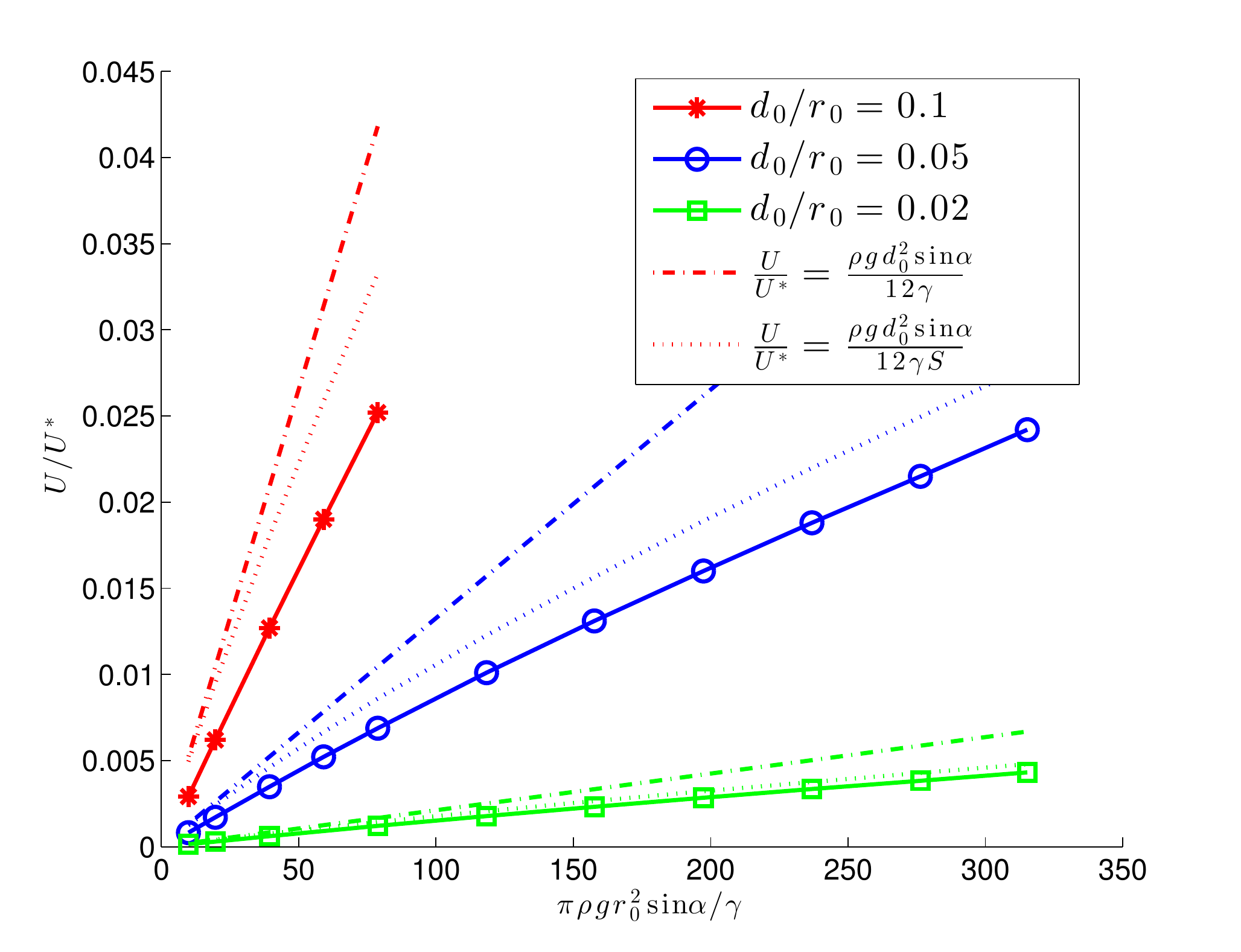}%height=4.5cm,
 %   \caption*{(a)Contact point position moves.}%加*可以去掉默认前缀，作为图片单独的说明
  %  \end{minipage}
    \caption{The relation between the rising velocity and the gravitational force of a bubble in a Hele-Shaw
cell.  The solid line is the result of the present theory.  The dash-dotted line is the result of 
Taylor and Saffman theory for circular bubble, Equ.~\eqref{e:velF2}. The dotted line is their result 
for elliptic bubble (Equ.~\eqref{e:velF1}) where the shape parameter $S$ \rev{is} calculated by Equ.~\eqref{e:steady2nonD} }%n张图片共享的说明
    \label{fig:CaForce}
\end{figure}

Fig.~\ref{fig:CaForce} shows the velocity plotted against the bond number. 
The solid lines indicate the velocity calculated by solving the equations~\eqref{e:steady1} and \eqref{e:steady2}, and 
the dotted and the dashed lines indicate  the velocity calculated by Equ.\eqref{e:velF1}
and Equ. \eqref{e:velF2} respectively.  It is seen that the simple circular model gives a 
reasonable estimate for the rising velocity.  The difference between the solid line and the dashed line
represents the effect of shape parameter. As we shall show in the following, the rising
bubble becomes flattened ($S>1$), and therefore  the rising velocity becomes smaller
than that of the circular bubble.  The difference between the dotted line and the solid
line represents the effect of Bretherton dissipation.  This term slows down
the rising velocity.  

\begin{figure}[!htp]
 \vspace{-0.5cm}
    \includegraphics[width=12cm]{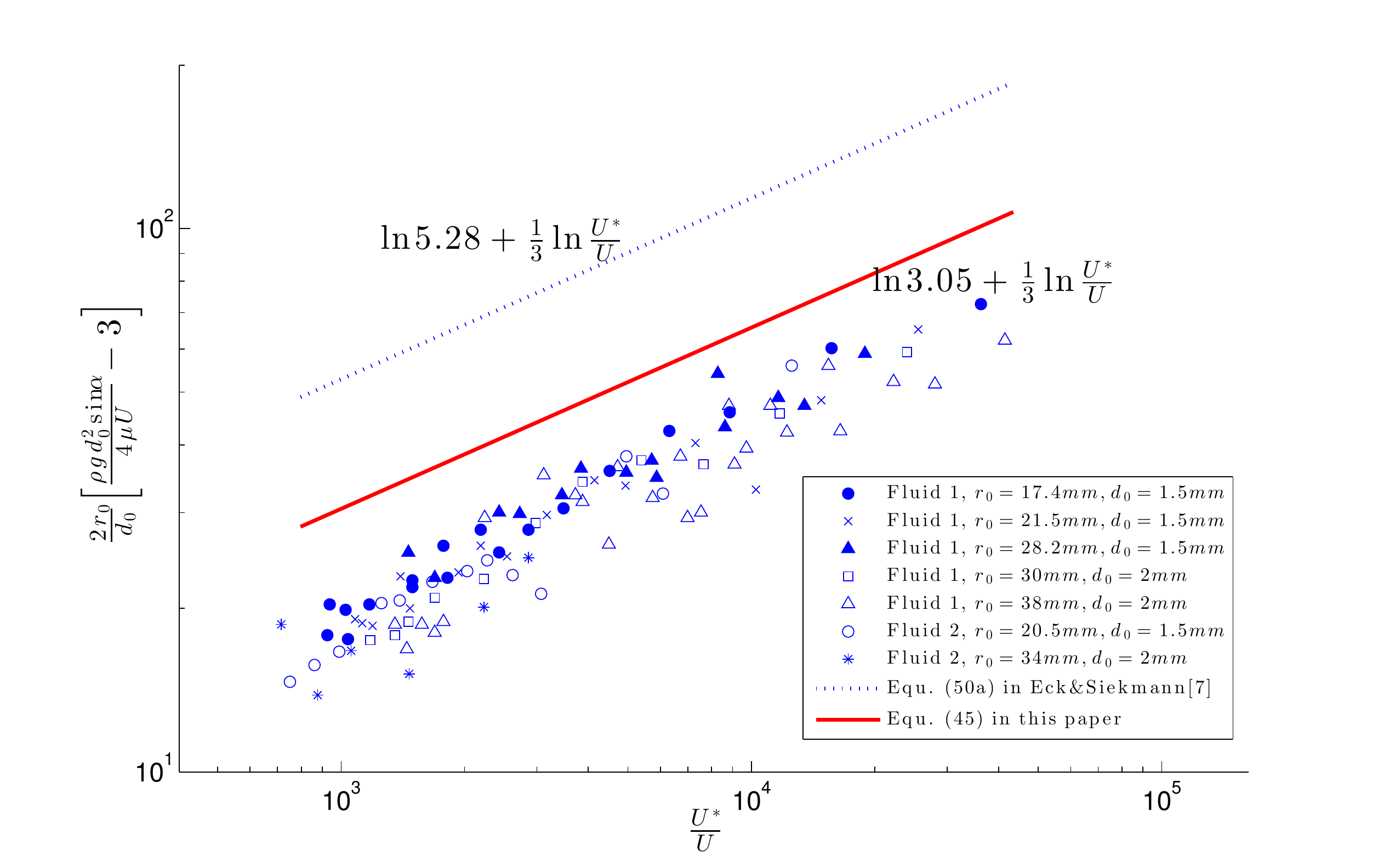}%height=4.5cm,
     \vspace{-0.5cm}
    \caption{\rev{Comparison between theories and experiment.
The dotted line is the result of Eck and Siekman (Equ. \eqref{e:compExp1})  and the solid line 
is our result (Equ.~\eqref{e:compExp}).  The marks are experimental data in  \cite{eck1978bubble}(Fluid 1: $60\%$ isopropanol, $40\%$ water; Fluid 2: $70\%$ glycerine, $18\%$ isopropanol, $12\%$ water).}} 
    \label{fig:comExp}
\end{figure}

The effect of Bretherton dissipation on the rising velocity was considered by 
Eck and Siekmann \cite{eck1978bubble}. \rev{ They obtained an expression
for the rising velocity of a circular bubble similar to Equ.~\eqref{e:steady1nonD}:
\begin{equation}
        3  + 5.28 \frac{d_0}{2 r_0} \Big(\frac{U^*}{U}\Big)^{1/3}=\frac{\rho g  d_0^2 \sin\alpha }{4 \mu U}.
\label{e:compExp1}
\end{equation}
For circular bubble, Equ.~\eqref{e:steady1nonD} gives the following result (with the use of 
$\tilde k_{11}(1)\approx 9.59$)
\begin{equation}
    3 +3.05\frac{d_0}{2 r_0} \Big(\frac{U^*}{U}\Big)^{1/3}=\frac{\rho g  d_0^2 \sin\alpha }{4 \mu U}.
   \label{e:compExp}
\end{equation}
The difference between Equ.~\eqref{e:compExp1} and Equ.~\eqref{e:compExp} are only in
coefficients.  They come from the difference in the estimation of the extra energy dissipation at the 
perimeter: Eck and Siekmann used the analysis of Fritz\cite{friz1965uber}, 
while we used the Park-Homsy boundary condition.  These results are compared in 
 Figure~\ref{fig:comExp} together with the experimental data obtained by
Eck and Siekmann \cite{eck1978bubble}.   (Notice that in the parameter range shown 
in  Figure~\ref{fig:comExp},  our result can be safely represented by the circular bubble 
since the capillary number $U/U^*$ is less than $10^{-3}$ (see Fig.~\ref{fig:CaForce} ). )   
Both results are qualitatively in agreement with experiments, but Equ.~\eqref{e:compExp} 
is closer to the experimental data, indicating that the Park-Homsy's boundary condition
(or Bretherton's analysis) is closer to reality.}

%\item 
\subsection{Shape of the rising bubble} 
\begin{figure}[!htp]
 %   \begin{minipage}[t]{0.33\linewidth}%设定图片下字的宽度，在此基础尽量满足图片的长宽
 %   \centering
    \includegraphics[width=9cm]{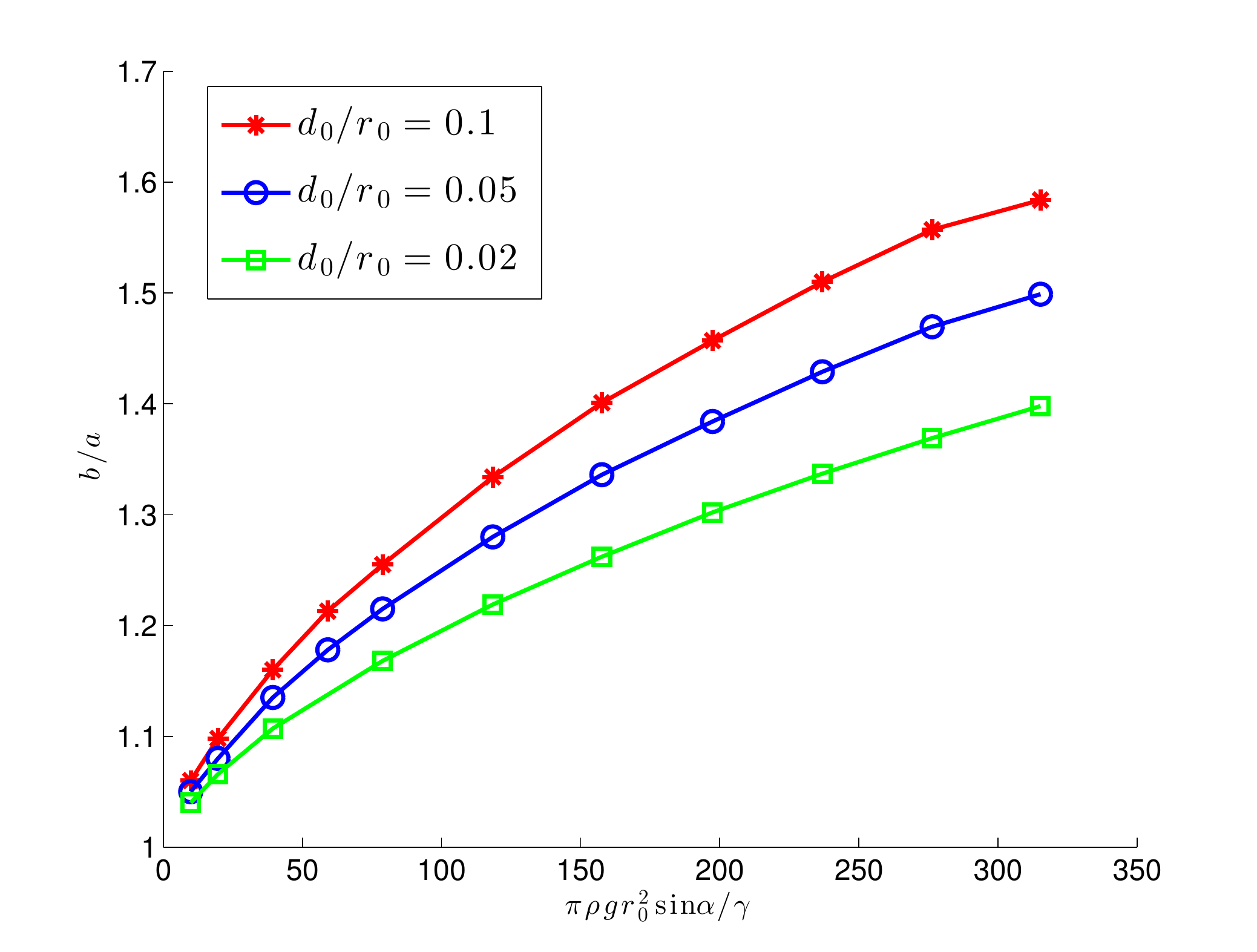}%height=4.5cm,
 %   \caption*{(a)Contact point position moves.}%加*可以去掉默认前缀，作为图片单独的说明
  %  \end{minipage}
    \caption{The relation between  the shape parameter $S=b/a$  and the gravitational force.}%n张图片共享的说明
    \label{fig:RatioBA_Force}
\end{figure}

Fig.~\ref{fig:RatioBA_Force} shows how the shape parameter $S=b/a$ changes with the 
Bond number $Bo_{\alpha}$.  The shape parameter $S$ is equal to 1 when $Bo_{\alpha}=0$.
As the bubble starts to rise, $S$ becomes larger than $1$ so the bubble is flattened.
It is important to note that this shape change is due to Bretherton dissipation. 
If the Bretherton dissipation is not considered, the circular bubble will rise keeping the 
circular shape as it was discussed  previously\cite{taylor1959note, tanveer1986effect}. 
The difference of the Bretherton dissipation in the advancing side and the receding side 
breaks the symmetry of the circular shape and causes the deformation of the bubble.

Fig.~\ref{fig:RatioBA_Force} shows that the shape change is larger in thick cell than in
thin cell. This is because the Bretherton effects become more significant in a thicker 
Hele-Shaw cell.

\begin{figure}[!htp]
 %   \begin{minipage}[t]{0.33\linewidth}%设定图片下字的宽度，在此基础尽量满足图片的长宽
 %   \centering
    \includegraphics[width=9cm]{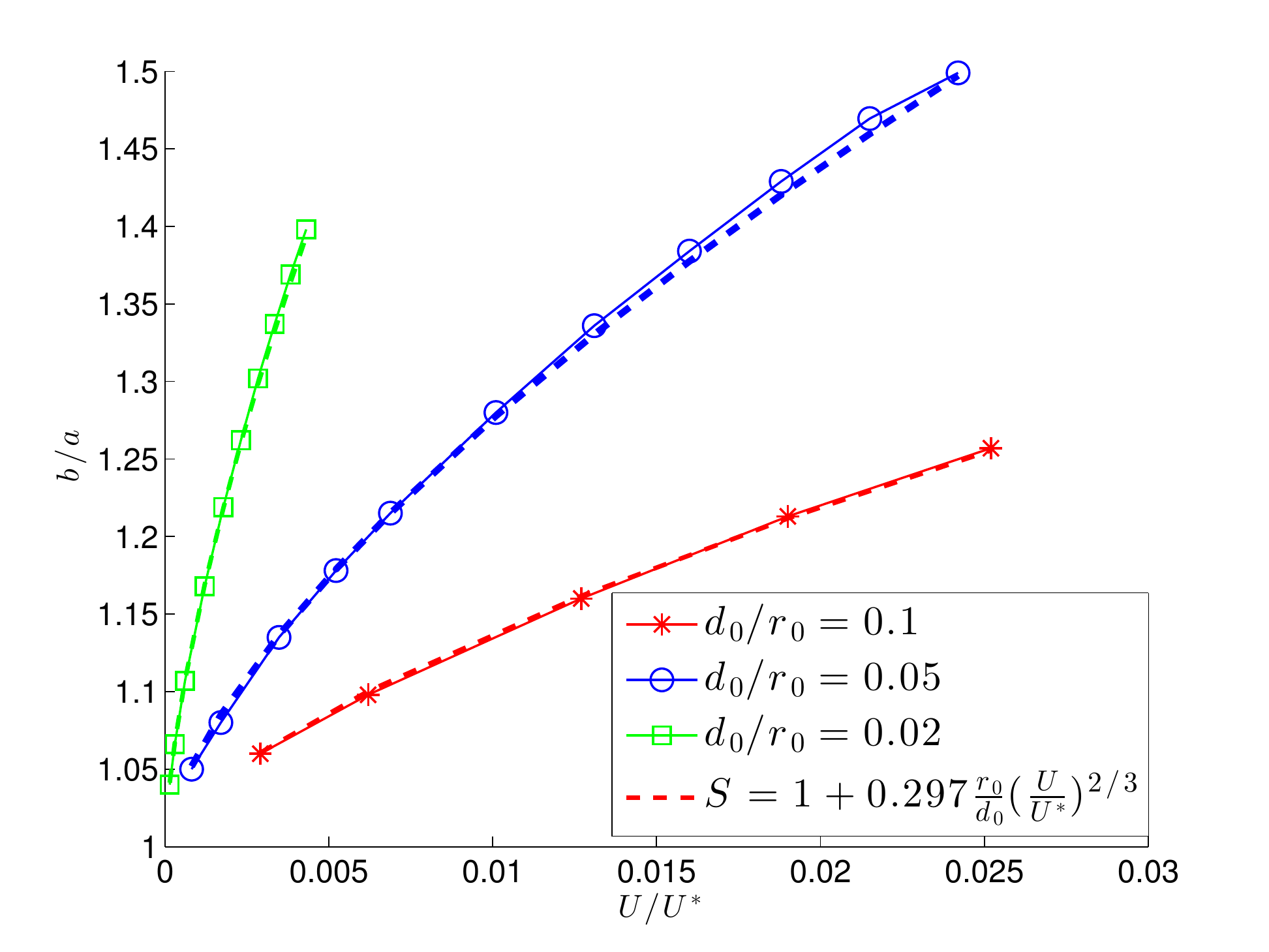}%height=4.5cm,
 %   \caption*{(a)Contact point position moves.}%加*可以去掉默认前缀，作为图片单独的说明
  %  \end{minipage}
    \caption{The relation between the shape parameter $S=b/a$ and the rising  velocity $U/U^*$ of 
the bubble.  The dashed line represents  Equ.~\eqref{e:steady2Appr1}.}%n张图片共享的说明
    \label{fig:RatioBA_Ca}
\end{figure}

Fig.~\ref{fig:RatioBA_Ca} shows the shape  parameter plotted against 
the rising velocity $U/U^*$.  The dashed line in Fig.~\ref{fig:RatioBA_Ca} 
represents the following simple equation
\begin{equation}\label{e:steady2Appr1}
      S=1+ 0.297 \frac{r_0}{ d_0} \cdot \left(\frac{U}{U^*} \right)^{2/3},
\end{equation}
This equation is obtained from Equ.~\eqref{e:steady2} by putting  $\tilde k_{12}$ equal to 1.1, the
asymptotic value of $\tilde k_{12}$  for large $S$ (see Fig. \ref{fig:Coef}).
Fig.~\ref{fig:RatioBA_Ca} shows that this simple relation reproduces the numerical results quite well.

\subsection{Effect of the bubble size.} 
\begin{figure}[!htp]
 %   \begin{minipage}[t]{0.33\linewidth}%设定图片下字的宽度，在此基础尽量满足图片的长宽
 %   \centering
    \includegraphics[width=9cm]{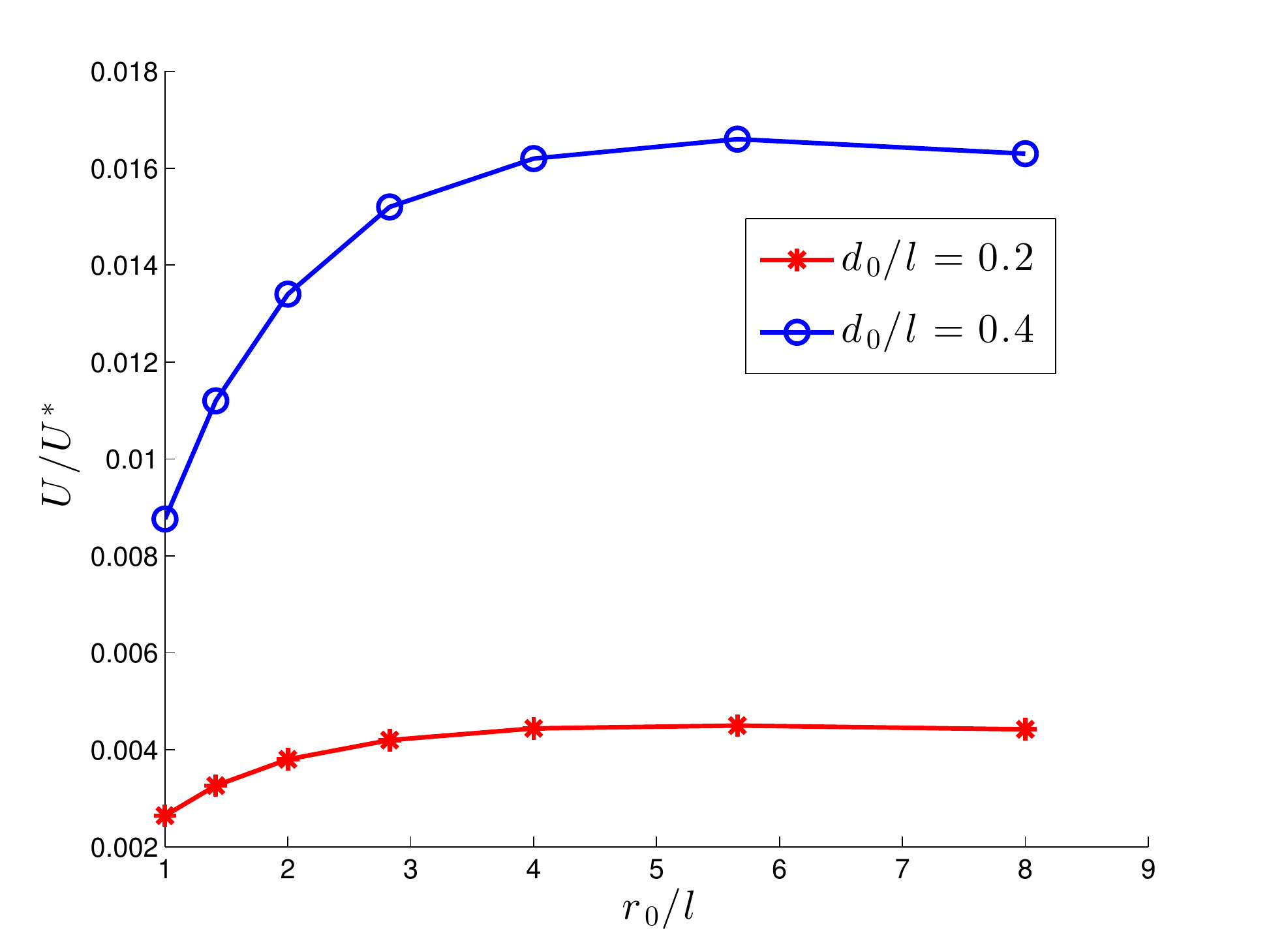}%height=4.5cm,
 %   \caption*{(a)Contact point position moves.}%加*可以去掉默认前缀，作为图片单独的说明
  %  \end{minipage}
    \caption{The relation between the rising velocity and \rev{the bubble size}.}%n张图片共享的说明
    \label{fig:Ca_R0Ex2}
\end{figure}

\begin{figure}[!htp]
 %   \begin{minipage}[t]{0.33\linewidth}%设定图片下字的宽度，在此基础尽量满足图片的长宽
 %   \centering
    \includegraphics[width=9cm]{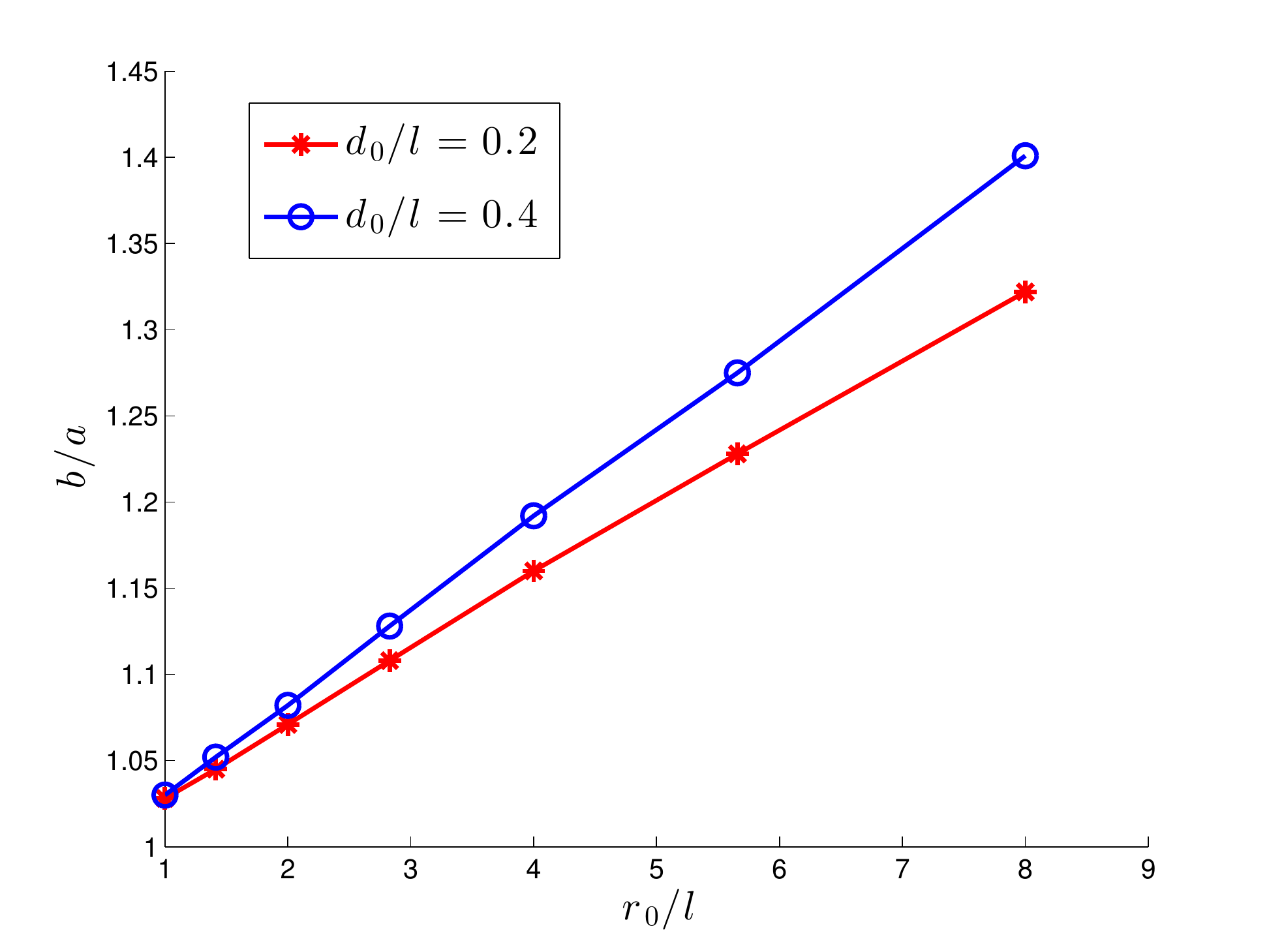}%height=4.5cm,
 %   \caption*{(a)Contact point position moves.}%加*可以去掉默认前缀，作为图片单独的说明
  %  \end{minipage}
    \caption{The relation between the shape parameter $S=b/a$ and \rev{the bubble size}.}%n张图片共享的说明
    \label{fig:Ratio_R0Ex2}
\end{figure}

We now study how the bubble size affects  the rising velocity
and its shape changes in a vertical cell($\alpha=\frac{\pi}{2}$).
We take the capillary length $l=\sqrt{\gamma/\rho g}$ as a reference length unit.
For given thickness of the Hele-Shaw cell, we change the  bubble size $r_0$, 
and solved Equ.\eqref{e:steady1} and \eqref{e:steady2}.  The results are shown in 
Fig.~\ref{fig:Ca_R0Ex2} and \ref{fig:Ratio_R0Ex2}. 

 Fig.~\ref{fig:Ca_R0Ex2} shows the rising velocity $U$ plotted against the bubble size $r_0$
for thick ($d_0/\ell =0.4$)  and thin($d_0/\ell =0.2$) cells.  It is seen that large bubbles rise with  
velocity independent of their size.  This is because the gravitational force and the frictional force 
are both proportional to the volume of the bubble in Hele-Shaw cell. The effect can be seen in the 
simple model (Equ.~\eqref{e:velF2}).   Small bubbles rise with size-dependent velocity, 
which is smaller than the asymptotic value.  This is due to the Bretherton dissipation: 
the Bretherton dissipation is proportional to the length of the perimeter and becomes significant
for smaller bubbles.

Careful inspection of Fig.~\ref{fig:Ca_R0Ex2} indicates that 
the rising velocity shows a small maximum as a function of $r_0$.  
The maximum arises from the two competing effects:  as the bubble size increases, the
effect of Bretherton dissipation decreases, while the effect of bulk dissipation increases due to the
flattening of the bubble.

%The existence of a maximal value for velocity can be understood from the equation~\eqref{e:steady1n} as follows. 
% When $r_0$ is large enough, the first term in the left hand side of \eqref{e:steady1n} becomes 
% dominant. We can ignore the Bretherton term and consider the equation~\eqref{e:velF1} instead. Noticing that  
% $Bo_{\alpha}= \frac{\rho g  r_0^2 \sin\alpha }{\gamma}=\frac{\rho g  r_0^2 }{\gamma}$ for $\alpha=\frac{\pi}{2}$, 
%  the equation is reduced to
%\begin{equation}\label{e:largeR0}
%\frac{U}{U^*} = \frac{\rho g  d_0^2 }{6\gamma S }=\frac{ 1}{6 S }\left(\frac{d_0}{l}\right)^2.
%\end{equation}
%It does not depends explicitly on $r_0$. If we choose $S=1$, 
%the equation gives a upper boundary of the velocity. 

 Fig.~\ref{fig:Ratio_R0Ex2} shows the shape parameter $S=b/a$ plotted against  $r_0/l$.
It is seen that $S$ increases linearly with $r_0$ and decreases with the increase of $d_0$.
Such behaviour can be understood from Equ.~\eqref{e:steady2Appr1}.  
%If we combine the equation \eqref{e:largeR0}
%with \eqref{e:steady2Appr}, we can obtain a formula for $S$,
%\begin{equation}
%\frac{2\tilde{k}_0 r_0}{d_0} (\frac{\rho g  d_0^2\sin\alpha}{6\gamma })^{2/3}=\frac{3\pi^2}{4}(S-1)S^{2/3}.
%\end{equation}
%which is valid for very large bubbles. 
%The  increase of the rising velocity becomes slower and slower and approaches to some constant value. 
%In the case, the increasing of the bubble size does not induce the increase of the velocity anymore.
%In the end, the shape of the bubble becomes more flat. 
%This makes the velocity decreases slightly from the maximal value, just as shown in the later stage of the velocity in Figure~\ref{fig:Ca_R0Ex2}.
%In particular, we show that for a given Hele-Shaw cell, the physical property of the fluid is fixed,

\section{Conclusions}
By using a variational principle, we have derived a simple evolution equation for the
shape change of  a rising bubble  in an \rev{infinitely} large  Hele-Shaw cell.
The equation explains the flattening of a rising bubble  observed in  
experiments\cite{kopf1988bubble,eck1978bubble}.  
Our analysis shows that the Bretherton dissipations 
is essential for the flattening. Without this term, the bubble would take a circular shape.
We gave quantitative prediction about the shape change and velocity of the bubble. They 
can be checked experimentally. 

In the present analysis, we have ignored the effect of the side boundary of a Hele-Shaw cell.
% That we can only get a circular bubble if without considering the Bretherton term or a fattened bubble with this effect. 
If the size of the Hele-Shaw cell is not large, the boundary effect makes the bubble elongated\cite{tanveer1986effect,tanveer1987stability,maxworthy1986bubble}. The competition of the Bretherton effect and
the boundary effect should be the reason for the complex shape changes of the bubble in a Hele-Shaw cell.
Indeed, Kopf-Sill and Homsy \rev{showed} that the flattening occurs only for bubbles relatively small \rev{compared} with the Hele-Shaw cell  \cite{kopf1988bubble}. Larger bubbles, on the other hand,  are elongated\cite{kopf1988bubble}. More theoretical
study is needed to quantify how the two effects together affect the shape change of an air bubble. This will be left for future work.

\section*{Acknowledgement}
This work was supported in part  by the National Key R\&D Program of China under Grant 2018YFB0704304 and Grant 2018YFB0704300(X.X.)
and by the National Natural Science Foundation of China under project nos. 11971469 (X.X), 11421110001(M.D.), 21774004(J.Z.) and 11771437(Y.D.).

%\appendix{\bf  Appendix. The dissipation in the transition region.}
%\numberwithin{equation}{section}
%\section{Calculations}
%\subsection{The expansions}
\section*{Appendix}
\setcounter{equation}{0}
\renewcommand{\theequation}{A\arabic{equation}}
\subsection{The Bretherton energy dissipations}
\begin{figure}[!htp]
 %   \begin{minipage}[t]{0.33\linewidth}%设定图片下字的宽度，在此基础尽量满足图片的长宽
 %   \centering
 \vspace{0.2cm}
    \includegraphics[width=7cm]{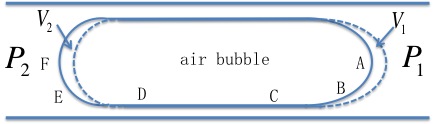}%height=4.5cm,
 %   \caption*{(a)Contact point position moves.}%加*可以去掉默认前缀，作为图片单独的说明
  %  \end{minipage}
%    \begin{minipage}[t]{0.33\linewidth}%需要几张添加即可，注意设定合适的linewidth
%    \centering
%    \includegraphics[height=4.5cm,width=6cm]{pics/k2eB.eps}
%    \caption*{(b)Contact angle changes.}
%    \end{minipage}
%    \begin{minipage}[t]{0.34\linewidth}%需要几张添加即可，注意设定合适的linewidth
%    \centering
%    \includegraphics[height=5cm,width=6cm]{pics/k2eC.eps}
%    \caption*{(c)The drop profiles changing by time.}
%    \end{minipage}
    \caption{A bubble between two plates.}%n张图片共享的说明
    \label{fig:Bretherton}
\end{figure}

In this subsection, we aim to compute the viscous energy dissipation in the vicinity of the boundary of a moving bubble in a Hele-Shaw cell.
Since the dissipation is related to  the classical analysis in Bretherton's paper \cite{Bretherton}, we call it a Bretherton energy dissipation term.
We consider a two dimensional problem. It is
 a long two-dimensional bubble in a channel between two solid boundaries as shown in Fig.~\ref{fig:Bretherton}.
The computations below are based on the previous analysis in \cite{Bretherton} and \cite{park1984two}.
% This can be done by using the analysis in Bretherton's classical paper \cite{Bretherton}. 

%This can be seen as the side view of a three-dimension bubble.
%This is slight different from that in \cite{Bretherton}, where a bubble in a circular tube is considered. 

We  analyse this problem by a generalized force balance argument. Suppose the fluid pressure in
the left side  is $P_2$ and that in the right side is $P_1$. We assume the bubble moves in  the right
direction. If the bubble moves for a short distance,  the liquid in the left part
changes with a volume $V_2$ and the right part changes with a volume $V_1$. Assume the air bubble is incompressible, 
then we have $V_1=V_2$. The free energy 
changes in this process are given by  $-(P_2 V_2-P_1 V_1)$. If the bubble moves
with a velocity $U$, the energy changing rate is given by 
$$\dot{ A}= -(P_2-P_1)d_0 U.
$$
Here $d_0$ is the thickness of the channel.
Then the driven force is given by $(P_2-P_1)d_0$. If we assume the energy dissipation function, which is half of the energy dissipation rate, is given by 
$$\Phi=\frac{1}{s}\xi U^s.$$
By  the Onsager principle, the driven force is balanced by the fiction force,
\begin{equation}\label{e:balance}
\xi U^{s-1}=(P_2-P_1)d_0.
\end{equation}
By the previous analysis in \cite{park1984two}, we  have the jump condition for pressures 
\begin{equation}
P-P_1\approx \frac{\gamma }{d_0/2}(1+\beta_1 (C_a)^{2/3}), \quad P- P_2 \approx \frac{\gamma }{d_0/2}(1-\beta_2(C_a)^{2/3})
\end{equation}
where $P$ is the pressure in the bubble, $C_a={\mu U}/{\gamma}$ is the capillary number, $\gamma$ is the surface tension, $\mu$ is
the viscosity of the fluid, $d_0$ is the distance between the two boundaries of the channel, $\beta_1 \approx 3.8$ and $\beta_2 \approx 1.9$ are two positive constants. We then have 
$$P_2-P_1\approx (\beta_1+\beta_2)\frac{2\gamma}{d_0}C_a^{2/3}=(\beta_1+\beta_2)\frac{2\gamma}{d_0}(\frac{\mu U}{\gamma})^{2/3}.$$
Combining it with the equation~\eqref{e:balance}, we have
\begin{equation*}
s= \frac{5}{3}, \qquad \xi=2(\beta_1+\beta_2){\gamma^{1/3}\mu^{2/3}}.
\end{equation*}
This gives a non-quadratic formula for the viscous energy dissipation in the vicinity of the bubble
\begin{equation}
\Phi=\frac{3\xi}{5} U^{5/3}=\frac{6(\beta_1+\beta_2)}{5}\mu (U^*)^{1/3} U^{5/3}   %C_a^{-1/3}\mu U^2.
\end{equation}
where $U^*={\gamma}/{\eta}$.
The above analysis can also done separately for the head and tail parts. Then we obtain the Bretherton energy dissipation terms
\begin{align}
\Phi_{head}=\frac{6\beta_1}{5}\mu (U^*)^{1/3} U^{5/3}\approx 4.56 \mu (U^*)^{1/3} U^{5/3},\label{e:bretherton0}\\
 \Phi_{tail}=\frac{6\beta_2}{5}\mu (U^*)^{1/3} U^{5/3} \approx 2.28\mu (U^*)^{1/3} U^{5/3},\label{e:bretherton1}
\end{align}
where we denote by $\Phi_{head}$ and $\Phi_{tail}$   the dissipation terms in the head and tail parts respectively.
\subsection{Solution of the Laplace equation in an infinite domain}
\setcounter{equation}{0}
\renewcommand{\theequation}{B\arabic{equation}}
When $r_0\ll L$, the boundary effect of the Hele-Shaw cell can be ignored.
In this case, the Laplace equation in Section 3 can be solved analytically.
We suppose $\hat{\Omega}^c=\mathbb{R}^2\setminus\hat{\Omega}$.
Then for $i=1,2$, we need solve
\begin{equation}
\left\{
\begin{array}{ll}
-\Delta \psi_i =0,&\qquad \hbox{in } \mathbb{R}^2\setminus\hat{\Omega},\\
\tilde{n}\cdot \nabla \psi_i = f_i(a/b,\hat{\theta}), & \qquad \hbox{on }\partial\hat{\Omega},\\
\psi_i\rightarrow const, & \qquad |\mathbf{r}| \hbox{ goes to infinity.}
\end{array}
\right.
\end{equation}
The  equation can be solved by a harmonic mapping method when $\hat{\Omega}$ is elliptic.

Introduce a harmonic mapping in
complex plane which maps a circle with radius $R=\hat{a}+\hat{b}$ to the ellipse $\hat{\Omega}$ with radii $\hat{a}=a/r_0$ 
and $\hat{b}=b/r_0$, 
\begin{equation}
z=\mathcal{F}(\zeta)=\frac{1}{2}(\zeta+\frac{c^2}{\zeta}),
\end{equation}
where $\hat c^2=\hat a^2-\hat b^2$. Here we choose the coordinate so that the radius $\hat a$ is in 
the $x$ direction. Its inverse mapping is given by $\zeta=\mathcal{F}^{-1}(z)=z+\sqrt{z^2-c^2}$.
The Laplace equation in an infinite domain outside a circular  can by solved explicitly. 
Actually, if a harmonic function $\phi$ outside a circle satisfies a Neumann boundary condition $\nabla \phi\cdot \mathbf{n}=g$
on the circle $|\zeta|=R$, then it can be computed from its boundary data as\cite{hitotumatu1954neumann}:
\begin{equation}\label{e:Sol_circle}
\phi(\zeta)=\frac{1}{\pi}\int_{|\tilde{\zeta}|=R}g(\tilde{\zeta})\ln \frac{1}{|\tilde{\zeta}-\zeta|} d\tilde{\zeta},
\end{equation}
for all $\zeta$ in the infinite domain. Using this formula, we could obtain the solution for $\psi_i$ as follows. 
%In the complex plane
%we used, the gravitational force $\mathbf{g}$ is along the real axis(negative direction). 
%Then the bubble will move along the real axis so that $U=(U,0)$.
%The
%complex number can also be written as $z=re^{i\theta}$ with $\theta$ is the angle from the positive real
%axis.
%An point $z=x+i y$ on the ellipse can be written as $x=a\cos\theta, y=b\sin\theta$. 
%The boundary condition can be rewritten as  
%\begin{equation}
%\mathbf{n}\cdot\mathbf{U}=\frac{\dot{a}b(2x^2-a^2)+Uabx}{a(a^4-c^2x^2)^{1/2}}
%=\frac{\dot{a}b\cos2\theta+Ub\cos\theta}{(a^2-c^2\cos^2(\theta))^{1/2}},
%\end{equation}
%and 
%\begin{equation}
%-\gamma\kappa-\rho \mathbf{g}\cdot\mathbf{r}=-\frac{2a^4b \gamma }{(a^4-c^2x^2)^{3/2}}-\rho g x
%=-\frac{2ab\gamma}{(a^2-c^2\cos^2\theta)^{3/2}}-\rho g a\cos\theta.
%\end{equation}
If we have $\partial_n \psi_i = f_i$ in $\partial_n\hat{\Omega}$, after the mapping, we have a harmonic  function 
$\phi_i(\zeta)=\psi_i(\mathcal{F} \zeta)$ outside a circle.  Direct computations give
$\partial_n \phi_i(\zeta) = \frac{\sqrt{ \hat{b}^2\cos^2\tilde{\theta}+\hat{a}^2\sin^2\tilde{\theta}}}{R}f_i(\frac{\hat{a}}{\hat{b}},\tilde{\theta})$, where $\zeta=R e^{i\tilde{\theta}}$.  
Using Equ.~\eqref{e:Sol_circle}, we obtain
\begin{equation}
\psi_i(z)=\phi_i (\tilde{r}e^{i \varphi})=\frac{1}{\pi}\int_{0}^{2\pi} f_i(\hat a/\hat b,\tilde{\theta})
\ln \frac{1}{|R e^{i\tilde{\theta}}-\tilde{r}e^{i \varphi}|}\sqrt{ \hat{b}^2\cos^2\tilde{\theta}+\hat{a}^2\sin^2\tilde{\theta}}d\tilde{\theta},
\end{equation}
where $ (\tilde{r}e^{i \varphi})=\mathcal{F}^{-1}z$.

Then we could compute the coefficients $k_{ij}$ as 
\begin{align*}
k_{ij}&=\int_{\partial \hat{\Omega}} f_i(s)\psi_j(s) ds=\int_0^{2\pi} f_i(\hat a/\hat b,\theta)\phi_j ({R} e^{i \theta}) \sqrt{\hat{b}^2\cos^2\theta+\hat{a}^2\sin^2\theta}d\theta\\
&=\frac{ 1}{\pi}\int_0^{2\pi} f_i(\hat a/\hat b,\theta) \int_{0}^{2\pi} f_j(\hat a/\hat b,\tilde{\theta})
\ln \frac{1}{|R e^{i\tilde{\theta}}-R e^{i \theta}|} \\ &
\qquad\qquad\qquad\qquad\qquad\qquad\qquad \times \sqrt{ \hat{b}^2\cos^2\tilde{\theta}+\hat{a}^2\sin^2\tilde{\theta}}d\tilde{\theta} \sqrt{\hat{b}^2\cos^2\theta+\hat{a}^2\sin^2\theta}d\theta \\
&=\frac{ 1}{\pi}\int_0^{2\pi} f_i(\hat a/\hat b,\theta) \int_{0}^{2\pi} f_j(\hat a/\hat b,\tilde{\theta})
\ln \frac{1}{\sqrt{(\cos\theta-\cos\tilde{\theta})^2+(\sin\theta-\sin\tilde{\theta})^2}}\\
&\qquad \qquad\qquad\qquad\qquad\qquad\qquad \times \sqrt{ \hat{b}^2\cos^2\tilde{\theta}+\hat{a}^2\sin^2\tilde{\theta}}d\tilde{\theta} \sqrt{\hat{b}^2\cos^2\theta+\hat{a}^2\sin^2\theta}d\theta\\
&=\frac{1}{\pi}\int_0^{2\pi}\int_{0}^{2\pi} f_i(\hat a/\hat b,\theta)  f_j(\hat a/\hat b,\tilde{\theta})
\ln \frac{1}{\sqrt{2(1-\cos(\tilde \theta-{\theta}))}}\\
&\qquad \qquad\qquad\qquad\qquad\qquad\qquad \times \sqrt{ \hat{b}^2\cos^2\tilde{\theta}+\hat{a}^2\sin^2\tilde{\theta}} \sqrt{\hat{b}^2\cos^2\theta+\hat{a}^2\sin^2\theta}d\tilde{\theta}d\theta.
\end{align*}
Direct computations gives
\begin{align*}
k_{11}
&=\frac{\hat{b}^2}{\pi}\int_0^{2\pi}\int_{0}^{2\pi} \cos(\theta) \cos(\tilde{\theta})
\ln \frac{1}{\sqrt{2(1-\cos(\tilde \theta-{\theta}))}}
d\tilde{\theta}d\theta\\
&=\frac{\hat{b}^2}{\pi}\int_0^{2\pi}\int_{0}^{2\pi} \cos(\theta) \cos(\tilde{\theta})
\ln \frac{1}{2|\sin((\tilde\theta-{\theta})/2)|}
d\tilde{\theta}d\theta\\
&=\frac{\hat{b}^2}{\pi}\int_0^{2\pi}\cos(\theta) \int_{0}^{2\pi} \cos(\hat{\theta}+\theta)
\ln \frac{1}{2|\sin(\hat{\theta}/2)|}
d\hat{\theta}d\theta\\
&=\frac{\hat{b}^2}{\pi}\int_0^{2\pi}\cos^2(\theta)d\theta \int_{0}^{2\pi} \cos(\hat{\theta})
\ln \frac{1}{2|\sin(\hat{\theta}/2)|}
d\hat{\theta}\\
&\quad -\frac{\hat{b}^2}{\pi}\int_0^{2\pi}\cos(\theta)\sin(\theta)d\theta \int_{0}^{2\pi} \sin(\hat{\theta})
\ln \frac{1}{2|\sin(\hat{\theta}/2)|}
d\hat{\theta}\\
&=\frac{\hat{b}^2}{\pi}\cdot\pi\cdot\pi +0=\frac{\pi {b}^2}{r_0^2},
\end{align*}
where we have used integration by part for  the term 
$$\int_{0}^{2\pi} \cos(\hat{\theta})
\ln \frac{1}{2|\sin(\hat{\theta}/2)|} d\hat{\theta}=2\int_{0}^{\pi} \cos(\hat{\theta})
\ln \frac{1}{2\sin(\hat{\theta}/2)} d\hat{\theta}=\pi.$$
Similarly,  we can obtain 
\begin{align*}
k_{12}=0, \qquad k_{22}=\frac{\pi {b}^2}{2r_0^2}.
\end{align*}
%where ${R}e^{i \theta}=\mathcal{F}^{-1}z=\mathcal{F}^{-1}(\hat{a}\cos\theta,\hat{b}\sin\theta)$.

{\bf Availability of Data.} The data that support the findings of this study are available from the corresponding author upon reasonable request.
{%\bibliographystyle{abbrv}%{elsarticle-harv}
\bibliography{Onsager}
}
\end{document}